\documentclass[useAMS,usegraphicx,usenatbib]{mn2e}

\title[On quasar host galaxies as tests of non-cosmological redshifts]{On quasar host galaxies as tests of non-cosmological redshifts} 
\author[E. Zackrisson]{E. Zackrisson\thanks{E-mail: ez@astro.uu.se}\\ Department of Astronomy and Space Physics, Box 515, S-75120 Uppsala, Sweden}
\begin{document}
\date{Accepted 2005 March 2. Received 2005 February 28; in original form 2004 July 16.}

\pagerange{\pageref{firstpage}--\pageref{lastpage}} \pubyear{2005}
\maketitle
\label{firstpage}

\begin{abstract}
Despite a general consensus in the astronomical community that all quasars are located at the distances implied by their redshifts, a number of observations still challenge this interpretation, possibly indicating that some subpopulation of quasars may harbour significant redshift components not related to the expansion of the universe. It has been suggested that these objects may have been ejected from local galaxies and are likely to evolve into new galaxies themselves. Here, a test of such exotic scenarios is proposed, based on the spectral energy distribution of the galaxies hosting quasars with suspected ejection origin. Provided that the time scales over which the ejected objects manifest themselves as quasars is short, one would in the framework of ejection scenarios expect to find either no quasar host galaxy, a pseudo-host consisting of gas ionized by the quasar, or a host galaxy consisting of young stars only. It is argued that the spectral energy distributions corresponding to the latter two options should differ significantly from those of most quasar host galaxies detected at low redshift so far, thus providing a potential test of the claimed existence of ejected quasars. A minimal implementation of this test, involving optical and near-IR broadband photometry, is suggested.
\end{abstract}
\maketitle

\section{Introduction}
The notion that quasars are not located at the distances implied by their redshifts, but rather situated nearby, dates all the way back to the discovery of the first quasars in the early 1960s (e.g. \citealt{Hoyle & Burbidge}). If quasars are to explain the observed X-ray background, a lower limit on the average quasar distance may however be inferred, indicating that only a small fraction of quasars can be truly local, i.e. located within a distance of 200 Mpc (\citealt{Burbidge a}). The discovery of galaxies and quasars located close in the sky and having very similar redshifts (\citealt{Stockton}), suggesting membership to the same group or cluster, also lends strong support to the cosmological interpretation of quasar redshifts. 

Despite a general consensus in the astronomical community that all quasar redshifts are in fact caused by the expansion of the universe, and hence good tracers of distance, a number of astronomers still defend the view that at least some subpopulation of quasars may harbour large components of non-cosmological redshift. Circumstantial evidence in favour of this hypothesis mainly comes in the form of apparent overdensities of high-redshift quasars in the field around nearby galaxies (e.g. \citealt{Zhu & Chu}), periodicities in the redshifts of such quasars (e.g. \citealt{Karlsson}; \citealt{Burbidge & Napier}), high-redshift quasars aligned along the minor axis of low-redshift active galaxies (e.g. \citealt{Arp c}; \citealt{Chu et al.}) and bridges of gas between low-redshift active galaxies and high-redshift quasars (\citealt{Sulentic & Arp}). \citet{Burbidge b} summarizes many of these findings.

Although such observations are often dismissed as due to a combination of gravitational lensing, chance projection and selection effects, it is still not clear whether this is enough to explain all of the reported curiosities (e.g. \citealt*{Hawkins et al.}; \citealt{Napier & Burbidge}; \citealt*{Benitez et al.}; \citealt{Gaztanaga}). 

One alternative, very controversial interpretation is that the quasars in these galaxy-quasar associations  have been ejected from active galaxies (e.g. \citealt{Arp c}). In this picture, it is argued that the ejected quasars -- through some still unknown mechanism -- evolve into normal galaxies or even clusters of galaxies (\citealt{Arp & Russell}) with the same redshifts as that of their `mother galaxies'. This scenario, although ignored by most astronomers, has proved notoriously difficult to falsify. Neither the presence of quasar absorption systems nor quasar host galaxies (see \citealt{Burbidge a} for a discussion) necessarily contradict the ejection hypothesis, as the absorbers may be interpreted as gas which due to its close proximity to the quasar itself has a non-cosmological redshift component, and the host galaxies as transition objects in the state of evolving from quasar to galaxy. It is therefore important to develop methods which have the potential to falsify the predictions of the ejection scenario and hopefully bring closure to the long-standing issue of non-cosmological redshifts. 

Although the existence of a quasar host system is not acknowledged as counterevidence to ejection by the proponents of non-cosmological redshifts, ejection scenarios may still impose a number of interesting constraints on the properties of the hosts. In this paper, a test based on the spectral energy distributions (SEDs) of quasar host galaxies is suggested. This method complements the proper motion test recently described by \citet{Popowski & Weinzierl} for the origin of quasar redshifts, but has the advantage of requiring much more modest observational facilities.

Section 2 describes the general principle of the host galaxy test and Section 3 a minimal implementation involving optical and near-IR broadband photometry only. In Section 4, a number of potential complications are discussed. Conclusions are summarized in Section 5.

\section{Quasar host galaxies as tests of quasar ejection scenarios}
Recently, \citet{Bell} presented an elaborate scenario of local quasar ejection, calibrated against quasars in the field around the active galaxy NGC 1068. This model has the virtue of being sufficiently detailed to be falsifiable. 

The \citet{Bell} model predicts that the ejected objects should manifest themselves as quasars for less than $10^8$ years before evolving into low-redshift objects and, supposedly, into normal galaxies.  By examining the quasars in the quasar-galaxy associations which are claimed to provide the best evidence for non-cosmological redshifts, one would therefore expect to find -- provided that these quasars really have been ejected in the proposed way -- either:
\begin{enumerate}
\item No host at all (because the transition from quasar to galaxy has yet to start, or because it may be unusually faint or small);

\item A host consisting only of gas ionized by the quasar. This gas may either have been accreted after ejection, ejected along with the quasar (as indicated by the claimed bridges) or produced (somehow) by the ejected object itself;

\item A host consisting only of gas and young stars with ages lower than $10^8$ years.

\end{enumerate}
These predictions stand in sharp contrast to the current understanding of quasar host galaxies. Despite early claims of detections of quasars with no host galaxy (\citealt*{Bahcall et al. a,Bahcall et al. b}), a general consensus has emerged that most quasars, if not all, are in fact hosted by galaxies. In cases where host galaxies are not found, the non-detections can typically be attributed to the faintness of the host relative to the active nucleus at the sampled wavelengths (e.g. \citealt*{Örndahl et al.}) or problems in subtracting the nuclear component due to the small angular size of the host, as in the case for quasars at high redshift. By targeting low-redshift quasars and going sufficiently deep in suitable passbands, one does however expect to detect the hosts unless they are unusually faint or small compared to those discovered so far. If the hosts of quasars located in peculiar galaxy-quasar associations systematically should turn out to have either of these characteristics, this would indicate that they indeed stem from a separate population.

\begin{figure}
\resizebox{\hsize}{!}{\includegraphics{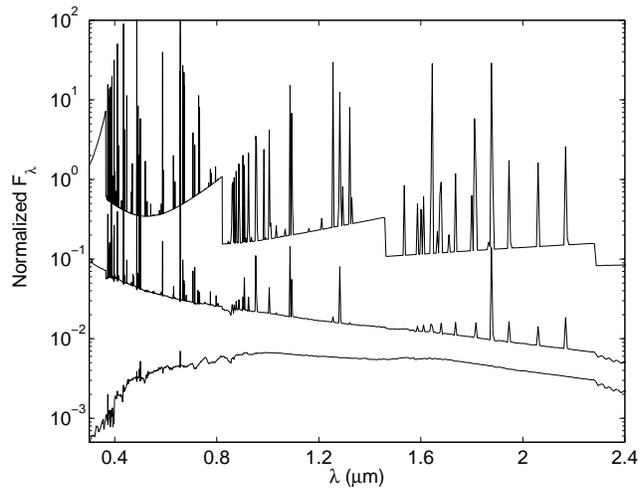}}
\caption[]{The optical to near-IR rest frame SED predicted for a a nebula photoionized by a quasar-like continuum located outside the observing beam (top), a 10 Myr old stellar population (middle) and a 5 Gyr old stellar population (bottom). The 10 Myr spectrum has been computed assuming a constant star formation rate considered appropriate for a young starburst, whereas the 5 Gyr spectrum has been computed assuming a decaying star formation rate with an e-folding decay rate of 1 Gyr considered appropriate for an early-type galaxy. In all cases, the gas is assumed to have solar metallicity. The vertical scaling of the spectra is arbitrary and only serves to display the various spectral features of the different spectra more clearly. The spectrum of a quasar-ionized nebula has been generated using the photoionization code Cloudy (\citealt{Ferland}), whereas the stellar population spectra have been produced by the \citet{Zackrisson et al.} spectral evolutionary model.}
\label{SEDfig}
\end{figure}

The majority of quasar host galaxies studied at low redshifts so far (e.g. \citealt*{Nolan et al.};\citealt*{Jahnke et al. a}) appear to have SEDs consistent with stellar populations at least a few Gyr old. As demonstrated in Fig.~\ref{SEDfig}, both the options (ii) and (iii) would produce SEDs of the host systems much bluer than this in the rest frame optical to near-IR wavelength range. Hence, many quasars appear not to have been ejected in the way suggested by \citet{Bell}. 
 
Since the quasars situated in the galaxy-quasar associations considered to provide the best cases for non-cosmological redshifts have not yet been purposely targeted for host galaxy investigations, the existence of a second population of quasar host galaxies with very different intrinsic properties can not be completely ruled out at the present time. By investigating whether the hosts of quasars located in such associations are consistent with either predictions (i), (ii) or (iii), this possibility can however be tested. Ideally, one would like to obtain high-dispersion spectra of the host galaxies all the way from the optical to the near-IR. Since this is rather expensive in terms of observing time, the possibility to carry out a more modest version of this test using optical/near-IR broadband photometry is explored in the following sections.  

\section{A minimal implementation of the quasar host galaxy test}
\subsection{The SEDs of stellar populations}
\begin{figure}
\resizebox{\hsize}{!}{\includegraphics{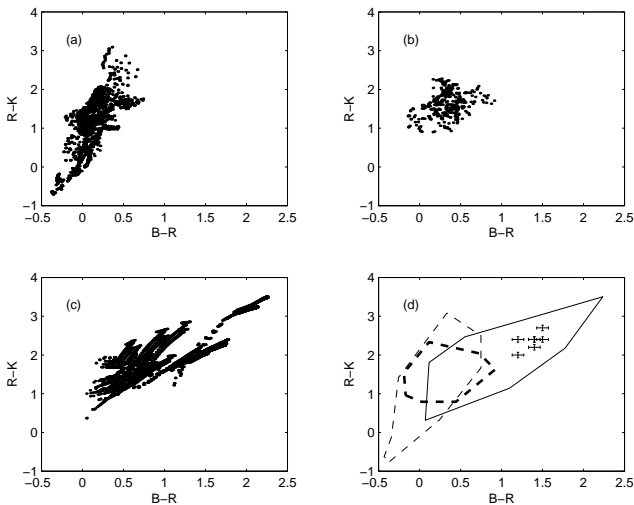}}
\caption[]{The position of stellar populations at ages $\leq 100$ Myr ({\bf a}), quasar-ionized nebulosities ({\bf b}) and stellar populations at ages $> 100$ Myr ({\bf c}) in diagrams of $B-R$ vs. $R-K$ at $z=0.10$. In {\bf d)}, the regions populated by {\bf a)}, {\bf b)} and {\bf c)} are outlined by polygons (thin dashed for ages $\leq 100$ Myr, thick dashed for quasar-ionized nebulosities, solid for ages $> 100$ Myr). Crosses indicate the colours and estimated errors of quasar host galaxies at approximately this redshift from \citet{Jahnke et al. a}. All these objects are located well within the region typical of old ($> 100$ Myr) stellar populations, indicating a non-ejection for these objects. In this diagram, two objects overlap due to identical colours.}
\label{colfig}
\end{figure}
To find a spectral diagnostic which can distinguish between ejection scenario prediction (iii) (young host) and the stellar populations typically found in normal host galaxies, a large number of synthetic stellar population SEDs covering the plausible range of ages, metallicities, star formation histories and initial mass functions (IMFs), have been computed using the spectral evolutionary model of \citet[][ hereafter Z01]{Zackrisson et al.}. In addition to predicting the SED of stellar populations over a wide range in age and metallicity, this model also includes the effects of both nebular continuum and line emission, which can be very important for young populations. The different values considered for each model parameter are listed in Table \ref{Z01grid}, and the resulting SED grid consists of every possible combination of these parameter values. 

Although the old and massive quasar hosts found so far at low redshifts are expected to be metal-rich, the young stellar population predicted by the ejection scenario may in principle have either low or high metallicity. In this case, the stars can be made from gas, supposedly of primordial chemical composition \citep*{Hoyle et al.}, produced by the ejected object, or from gas of any metallicity accreted after ejection or ejected along with the quasar from the mother galaxy. For this reason, stellar metallicities ranging from low ($Z=0.001$) to very high ($Z=0.040$) are considered. For simplicity, the metallicity of the gas is assumed to be the same as that of the stars.

The Salpeter initial mass function (IMF) is currently believed to be a reasonably accurate description of the IMF for stars more massive than $\approx 0.5 \ M_\odot$ (\citealt{Kroupa}). The Tully--Fisher relation also indicates that it may be fairly universal (e.g. \citealt{Bell & de Jong}). Here, a single-valued power-law ($dN/dM \propto M^{-\alpha}$) with a slope of $\alpha=2.35\pm 0.5$ is adopted over the standard mass range 0.08--120 $M_\odot$. Even though a single-valued IMF in this range may give an overestimate of the fraction of $<0.5\ M_\odot$ stars, such low-mass stars have very little impact on the predicted broadband colours and therefore do not significantly affect the present study.

Here, three different star formation histories are considered: A short burst lasting for 10 Myr after which the star formation rate (SFR) drops to zero; a constant SFR from 0 to 15 Gyr; and a SFR which increases exponentially over time, here parametrized as a negative e-folding decay rate of $\tau=-2$ Gyr, where $\mathrm{SFR(t)}\propto \exp (-t/\tau)$. Although these star formation scenarios cannot be expected to reproduce the SEDs of every galaxy type along the Hubble sequence, they are considered to cover the plausible range in terms of how fast or slow the SEDs of realistic stellar populations may evolve from blue to red as they age. Normal galaxies are typically assumed to have star formation rates which decrease over time, and are therefore located somewhere in between the short burst and the constant star formation scenario. It is not completely clear if galaxies exist which have star formation histories which increase over cosmological time scales, although such scenarios have been advocated for both irregular (\citealt*{Gallagher et al.}; \citealt{Sandage}) and low surface brightness galaxies (\citealt{Boissier et al.}). It may be argued that such scenarios could also arise in the hosts predicted in the ejection scenario, depending on the details of the gas production mechanism in this case.

Even though the escape fraction of Lyman continuum photons from normal galaxies is believed to be very low, a significant leakage may be imagined in the case of very violent starbursts, which could possibly be what the young hosts predicted in the ejection scenario would manifest themselves as. For this reason, both gas covering factor values of 1.0 (no leakage) and 0.0 (complete leakage) are tested, to cover both extremes.

A number of other gas parameters such as the hydrogen density and the filling factor also affect the predicted SED, although to a much smaller degree than the covering factor (see Z01 for a demonstration). Here, a filling factor of 0.01 and a hydrogen number density of 100 cm$^{-3}$, which are values typical for the interstellar medium in galaxies, are assumed.

\begin{table}
\caption[]{The grid of Z01 SEDs for stellar populations. All evolutionary sequences assume $Z_\mathrm{gas}=Z_\mathrm{stars}$, a total gas mass of $10^{10} \ M_\odot$ available for star formation and the IMF to be valid throughout the stellar mass range 0.08--120 $M_\odot$. Gas parameters typical of the interstellar medium in galaxies are adopted: A filling factor of 0.01 and a hydrogen number density of 100 cm$^{-3}$. The grid consists of evolutionary sequences for all possible combinations of the parameter values listed below. For each combination, SEDs are predicted at ages from 0.5 Myr to 15 Gyr in small time steps.}
\begin{flushleft}
\begin{tabular}{ll}
\hline
Metallicity, $Z$: & 0.001, 0.004, 0.008, 0.020, 0.040\cr
IMF, $\alpha$: & 1.85, 2.35, 2.85\cr 
SFH: & c$10^7$, c$15\times 10^9$, e$-2\times 10^9$\cr 
Covering factor:&  0.0, 1.0\cr 
\hline
\end{tabular}\\
IMF: $dN/dM \propto M^{-\alpha}$\\
SFH=Star formation history.\\
c=Constant star formation rate during the subsequent number of years.\\
e=Exponentially declining/increasing star formation rate  with an e-folding decay rate equal to the subsequent number of years. \\
\end{flushleft}
\label{Z01grid}
\end{table}

\begin{table}
\caption[]{The grid of quasar-ionized nebulosities, calculated using the photoionization model Cloudy (\citealt{Ferland}). The grid consists of all possible combinations of the parameter values listed below. See main text for details.}
\begin{flushleft}
\begin{tabular}{ll}
\hline
Continuum type: & 1, 2\cr
Metallicity, $Z$: & 0.000, 0.001, 0.004, 0.020, 0.050, 0.1000\cr
$M_\mathrm{V}$: & $-10$, $-15$, $-20$\cr 
Inner radius (pc): & 0.01, 1\cr
$\log n(\mathrm{H})$ (cm$^{-3}$): & 0, 2, 4\cr
Filling factor:  & 0.001, 1\cr
\hline
\end{tabular}\\
\end{flushleft}
\label{AGNnebgrid}
\end{table}

\subsection{The SEDs of quasar-ionized nebulae} 
To find a spectral diagnostic which can distinguish between ejection scenario prediction (ii) (gas ionized by the quasar) and the quasar hosts discovered so far, synthetic SEDs for nebulae photoionized by quasars located outside the observed beam have been generated using the photoionization code Cloudy version 90.05 (\citealt{Ferland}). All calculations assume a spherical, radiation-bounded and homogeneous nebula of constant density. The different values considered for each model parameter are listed in Table \ref{AGNnebgrid}, and the grid of quasar-ionized nebulae consists of every possible combination of such parameter values. For a small number of parameter combinations, Cloudy did however fail to produce physically acceptable ionization structures. For this reason, a total of 19 parameter combinations (out of the 432 allowed by Table \ref{AGNnebgrid}) were excluded from further analysis.

Due to the difficulties in observing the intrinsic SED of quasars prior to interaction with the surrounding interstellar medium, the continuum shape of the central light source most appropriate for model calculations is still uncertain. Here, two different quasar SEDs will be tested as ionization sources. Firstly, the multi-component continuum generated by the Cloudy AGN-command using default settings, and secondly a single power-law continuum ($f_\nu \propto \nu^\alpha$) with slope $\alpha=-1.0$ from the infrared through X-rays. In Table \ref{AGNnebgrid} these are referred to as type 1 and 2, respectively.

As previously argued, the metallicity of the gas surrounding the quasar may be either high or low, depending on its origin. Here, values of $Z=0.000$, 0.001, 0.004, 0.020, 0.050 and 0.100 are considered. Scaled solar abundances are assumed in all cases except at zero metallicity, where primordial abundances of both helium and heavier elements are adopted. 

If the quasars truly are located nearby, their luminosities must be rather low. Here, absolute V magnitudes of $M_\mathrm{V}=-10$, $-15$ and $-20$ are used, which covers the range derived by \citet{Burbidge a}.

Given the low luminosities of the quasars, the gas density must be rather low, on the same order as that typically observed in galaxies, in order to produce a Str\"omgren sphere large enough to provide a host galaxy detection. Here, hydrogen number densities (cm$^{-3}$) of $\log n(\mathrm{H})=0$, 2 and 4, as well as gas filling factors of 0.001 and 1.0 are considered.

The inner radius of the nebula, $r_\mathrm{in}$, together with $M_\mathrm{V}$, $n(\mathrm{H})$ and the shape of the ionizing continuum, define the ionization parameter $U$ of the nebula: $U=Q/(4\pi r_\mathrm{in}^2 n(\mathrm{H}) \mathrm{c})$, where Q is the number of hydrogen-ionizing photons emitted per second (set by the continuum shape and  $M_\mathrm{V}$) and c is the speed of light. If the gas is produced continuously by the quasar, we may expect the inner radius to be small. Here, values of $r_\mathrm{in}=0.01$ and 1 pc are used. Using the adopted range of all parameter values, the model grid covers a ionization parameter range of $-2.5 \leq \log U \leq 9.5$.

\subsection{Colour criteria}
In order to obtain information on the shape of the quasar host galaxy SED, broadband photometry in two filters must minimally be carried out. Here, a search for combinations of broadband filters which can efficiently distinguish quasar host galaxies from the predictions of the ejection scenario is presented.

For all the considered spectral energy distributions produced by stellar populations (Table~\ref{Z01grid}) or quasar-ionized nebulosities (Table~\ref{AGNnebgrid}), fluxes in $B$, $V$, $R$, $I$, $J$, $H$ and $K$ have been calculated for redshifts $z=0.0$--0.7 in steps of 0.05. For $B$, $V$, $J$, $H$ and $K$, Johnson filters profiles have been used, whereas for $R$ and $I$, ESO-filters 608 and 610 have been adopted, providing a reasonable approximation to the original Cousins-system. Quasars at $z>0.7$ are not considered here, because of current difficulties in subtracting the light from the nuclear component associated with the small angular size expected even for normal host galaxies at these distances. This could give rise to spurious non-detections, even if the host galaxies are not unusually faint compared to their active nuclei.

For the proposed test, the ideal filter combination would be one in which the considered stellar populations evolve reasonably monotonically towards the red with increasing age irrespective of redshift, and in which the possible colour range of the quasar-ionized nebulosities do not overlap with the colours of stellar populations at ages above 100 Myr. As it turns out, there is no single colour for which this criterion is fulfilled over the entire redshift range. The colours of short burst and high-metallicity populations typically feature a very red peak lasting a few times $10^7$ yr shortly after the end of active star formation, making high and low ages difficult to disentangle using a single colour only. The wide range of colours predicted for the quasar-ionized nebulosities, in addition to their rapid colour evolution with redshift as the strong emission lines and discontinuities evident from Fig.~\ref{SEDfig} shift in and out of filter boundaries, also severely complicates the analysis. 

\subsubsection{$z\la 0.35$}
At $z\la 0.35$, no single colour appears particularly useful over more than a very limited redshift range. Instead, a combination of $B$, $R$ and $K$ is suggested for $z \la 0.20$, followed by $B$, $V$ and $R$ at $0.20 \la z \la 0.35$. 

\begin{figure*}
\centering
\includegraphics[width=17cm]{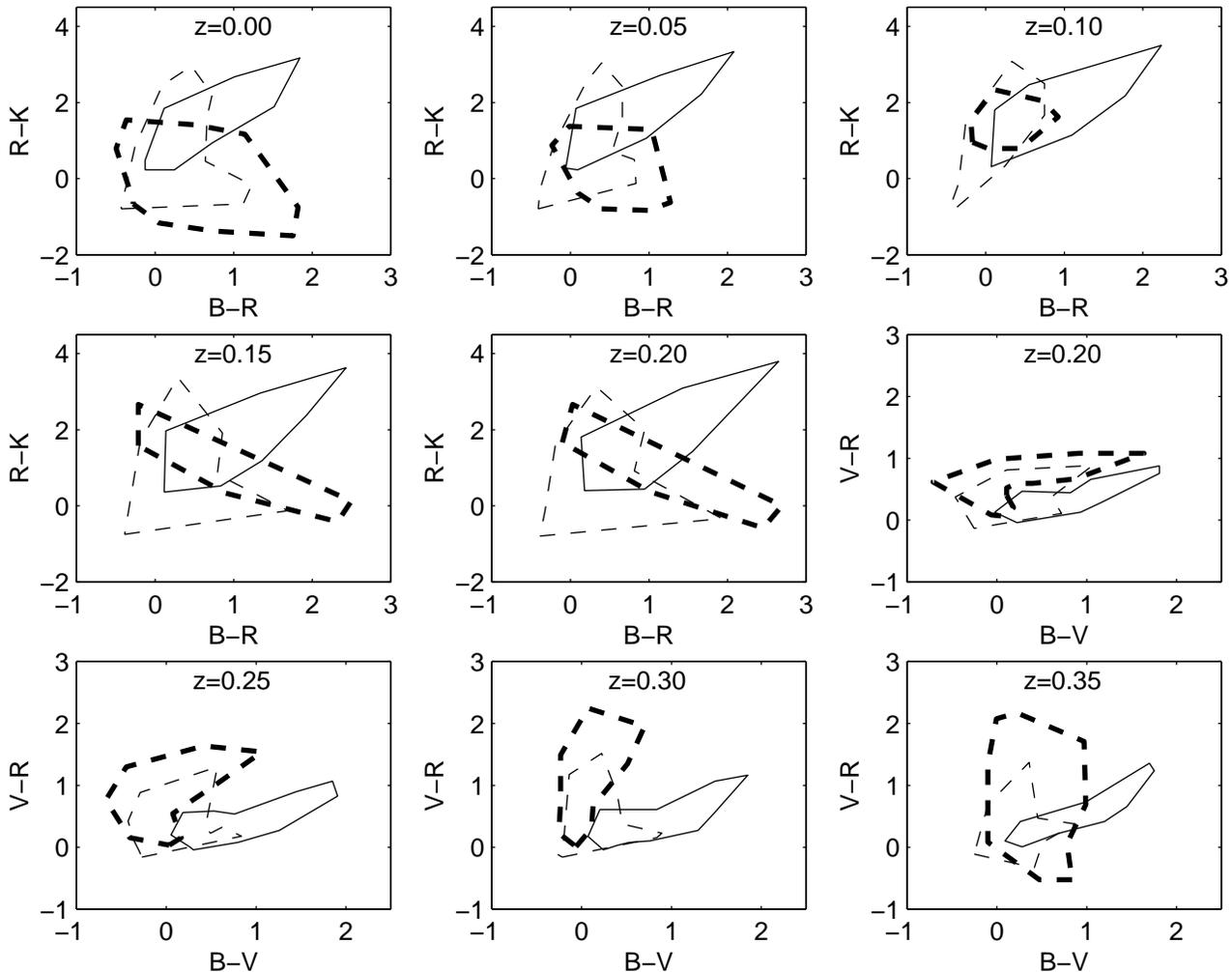}
\caption[]{The regions inhabited by  young ($\leq 100$ Myr) stellar populations (thin dashed), old ($> 100$ Myr) stellar populations (thin solid) and quasar-ionized nebulosities (thick dashed) in diagrams of $B-R$ vs. $R-K$ at $z\leq 0.20$ and $B-V$ vs. $V-R$ at $0.20 \leq z \leq 0.35$.}
\label{colfig2}
\end{figure*}

In Fig.~\ref{colfig}, the situation at $z=0.10$ is illustrated. Young ($\leq 100$ Myr) stellar populations, quasar-ionized nebulosities and old ($> 100$ Myr) stellar populations populate different regions in the diagram of $B-R$ vs. $R-K$, with only a small overlap between the two former and the latter. The position of seven quasar host galaxies from \citet{Jahnke et al. a} with $0.075 \leq z \leq 0.125$ are clearly located in the region uniquely corresponding to high ages, indicating a non-ejection origin for these objects. At this specific redshift, $B-R$ can by itself be used for a fairly decisive test, but due to the rapid redshift evolution of the quasar-nebulosities in this diagram, this is not the typical case. In Fig.~\ref{colfig2}, the redshift evolution of the three regions in $B-R$ vs. $R-K$ (at $z\leq 0.20$) or $B-V$ vs. $V-R$ (at $0.20 \leq z \leq 0.35$) are plotted. The young stellar populations, but in particular the quasar-ionized nebulosities, shift significantly between adjacent redshift bins, whereas the old stellar populations (lacking prominent emission lines) remain almost inert. 

A certain overlap between the region inhabited by old populations (delimited by solid line) and those of young populations (thin dashed) or quasar-ionized nebulosities (thick dashed) is unavoidable, but for the filter combinations used here typically small. Hence, it is likely that a firm outcome of the proposed test can be achieved using the suggested filters.

When determining the extent of the regions populated by old ($>100$ Myr) populations (solid lines) in Fig.~\ref{colfig} and~\ref{colfig2}, only ages lower than the age of the universe at each particular redshift have been included, assuming a $\Omega_\mathrm{M}=0.3$, $\Omega_\Lambda=0.7$ cosmology with $H_0=72 \  \mathrm{km \ s^{-1} Mpc^{-1}}$.

\subsubsection{$0.35 \la z \la 0.70$}
In the interval $0.35 \la z \la 0.70$, single colours are sufficient to separate young stellar populations and quasar-ionized nebulae from old stellar populations. At $0.35 \la z \la 0.4$, $V-I$ may be used, followed by $V-J$ at $0.40 \la z \la 0.70$.

\begin{figure*}
\centering
\includegraphics[width=17cm]{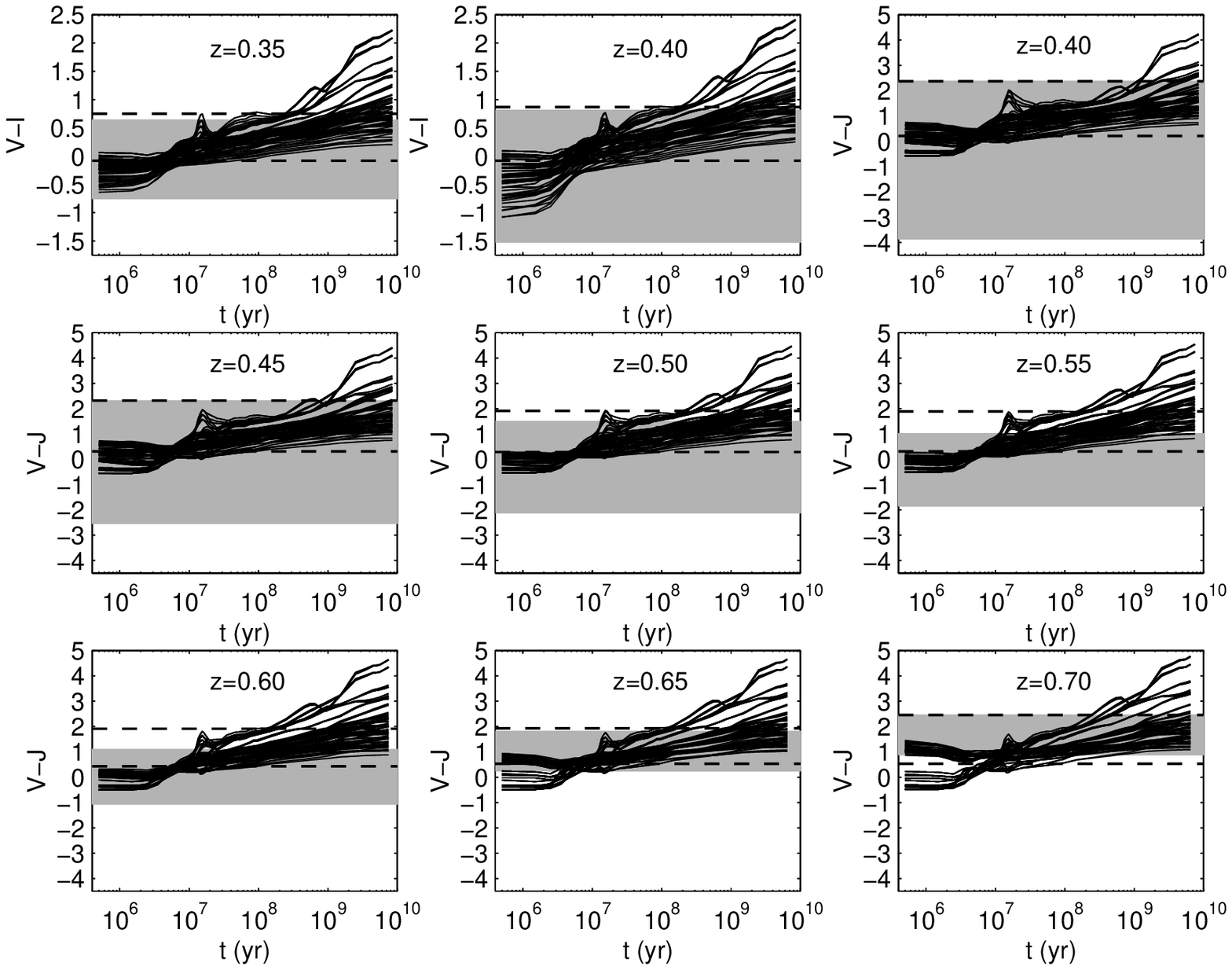}
\caption[]{The evolution of $V-I$ at $z=0.35$--0.4 and $V-J$ at $z=0.4$--0.70 for stellar populations with parameters defined by Table~\ref{Z01grid} (solid lines). Grey regions indicate the span of colours predicted for quasar-ionized nebulosities. Dashed lines indicate the boundaries of the region inside which the test will yield no firm outcome. Above the upper dashed lines, only stellar populations with ages above 100 Myr are found, whereas below the lower ones only young stellar populations or quasar-ionized nebulosities reside. The evolutionary tracks of the stellar populations terminate at the age of the universe at each $z$, assuming a $\Omega_\mathrm{M}=0.3$, $\Omega_\Lambda=0.7$ cosmology with $H_0=72 \  \mathrm{km \ s^{-1} Mpc^{-1}}$.}
\label{colfig3}
\end{figure*}

In Fig.~\ref{colfig3}, the evolution of stellar populations (solid lines) in $V-I$ and $V-J$ is plotted as a function of age, along with the colour range covered by quasar-ionized nebulosities (grey region), at redshifts of $0.35 \leq z \leq 0.40$ and $0.40 \leq z \leq 0.70$ respectively. Even though stronger constraints can be imposed for specific redshifts, a $V-I>0.9$ at $0.35 \leq z \leq 0.40$ would be inconsistent with ages lower than 100 Myr or quasar-ionized nebulae, and hence reject the ejection scenario. A colour of $V-I<-0.1$ would on the other hand be inconsistent with an age higher than 100 Myr, but consistent with either a pristine population or a quasar-ionized nebula, and thereby be in favour of an ejection origin. In the intermediate range, $-0.1 \leq V-I \leq 0.9$, various degeneracies prevent any firm outcome of the test. Similarly, $V-J>2.5$ at $z=0.40$--0.70 would be inconsistent with the ejection scenario and $V-J<0.2$ inconsistent with ages higher than 100 Myr. 

The reddest colours at ages $\geq 100$ Myr in Fig.~\ref{colfig3} are produced by short burst scenarios (c10$^7$ in Table \ref{Z01grid}). Among these, high metallicities are responsible for the reddest colours, and low metallicities for those slightly bluer. At high ages, the constant SFR scenarios (c$15\times10^9$) typically fall close to, or slightly below the upper dashed line which delimits the zone of ambiguity. As the galaxies along the Hubble sequence are believed to be characterized by star formation episodes less extended than the constant SFR scenario (but more so than the short burst), most normal galaxies are expected to fall above the dashed line in these diagrams. The variations in IMF slope are, within the explored range, only of minor importance. 

The dispersion in colour among the different stellar populations defined in Table \ref{Z01grid} is typically smaller at ages below $100$ Myr. The prominent red peak seen at all redshifts at $t\sim 10$ Myr is caused by high-metallicity ($Z>0.080$) populations. At the lowest ages, a clear separation in colour develop at certain redshifts, most notably at $z=0.70$, between stellar populations with and without nebular emission (covering factor 1.0 and 0.0, respectively). In this case, the nebular emission acts to make $V-J$ redder.

\section{Discussion}
In the colour-colour and single-colour diagrams discussed in section 3, the ejection scenario typically predicts the host galaxies to be bluer than the old host galaxies detected so far. If the quasars located in the galaxy-quasar associations which are claimed to provide the best evidence for non-cosmological redshifts turn out to have very blue colours, whereas other hosts at the similar redshift do not, this would indicate that they do in fact belong to a separate population. 

By targeting the brightest quasars of galaxy-quasar associations suspected of non-cosmological redshifts, it should be possible to carry out the required optical and near-IR photometry at a small telescope, as demonstrated by the host galaxy study carried out by \citet{Örndahl et al.} at similar redshifts with the 2.5 m Nordic Optical Telescope. To cover the $\approx 6$ targets with suitable redshifts among the quasars associated with NGC 1068, on which the model of \citet{Bell} was based, only a couple of nights of observing time would be required. This minimal implementation therefore has the obvious advantage of requiring much more modest facilities than the test proposed by \citet{Popowski & Weinzierl}, which requires VLT, Keck, HST or preferably planned satellites like GAIA and SIM.

Although the host galaxy modelling presented in section 3 has been fairly generous in terms of the curious host properties considered, a number of additional complexities still require consideration. 

\subsection{Recent star formation}
In order to be able to distinguish the quasar host galaxies predicted by the \citet{Bell} ejection scenario from those of the general population, the chosen diagnostics must of course be able to separate the two. This is straightforward as long as the probed wavelength region of the SED in normal hosts contains an obvious signature of an old stellar population -- as in the case of the \citet{Jahnke et al. a} objects in Fig.~\ref{colfig}. If a sufficiently powerful starburst is ignited in an old galaxy, the young population could however dominate part of the spectrum and lead to potential confusion with the predictions of the non-cosmological redshift model. Indeed, certain quasar host galaxies have been found to contain prominent young stellar populations \citep[e.g.][]{Canalizo & Stockton}, which in optical filters could give the appearance of a truly young host. However, in order for this to be a serious problem, starbursts would have to be very powerful and very common among quasar host galaxies. If, for instance, a young stellar population (Age 0--30 Myr, $Z=0.020$) is added to an old host (Age 10 Gyr, $\tau=5$ Gyr, $Z=0.020$) at a redshift of $z=0.10$ (corresponding to Fig.~\ref{colfig}) the young component would have to contribute $>50$\% to the V-band flux ($\sim 5000$ \AA{} in rest frame) to shift the $B-R$ colour of the galaxy into the colour span predicted for a $\leq 10^8$ yr object. Even though a few of the objects analyzed by \citet{Canalizo & Stockton}, which -- being transition objects between classical quasars and ultraluminous galaxies -- are likely to be biased in favour of recent star formation, may contain sufficiently young and prominent starburst populations to meet this criterion, this does not appear to be the typical case in the general population. Hence, although the test could be ambiguous for a single quasar host galaxy, the conclusions would be robust when applied to a larger sample.

\subsection{Ejection of old stars}
Is it possible that red colours of quasar host galaxies could be produced by old stars ejected along with the quasar from their parent galaxy? It is somewhat difficult to see how a large, collisionless system of stars could be accelerated to $\sim 0.1$c (\citealt{Bell}) without leaving a prominent tail (which supposedly would turn up in the host galaxy image). Even if this would occur, the proposed test could however strongly constrain the parameters involved in the scenario of evolution from quasar to normal galaxy. In order to successfully make this transition, it is not sufficient for the quasar to simply fade away. The luminosity of the underlying system must also significantly increase, supposedly through creation of matter and subsequent star formation. Provided that several hosts around possibly ejected quasars turn out to have SEDs consistent with old stars only, one would then -- given the short lifetime of the quasars derived by \citet{Bell} -- have to assume a scenario in which the quasar first fades from sight before the onset of significant star formation. 

\subsection{Dust reddening}
Even though the host galaxies of recently ejected quasars are expected to be very blue, red colours could arise even in the hosts of ejected objects if dust extinction is significant. If one assumes the existence of two separate populations of quasars, one featuring very blue intrinsic colours and the other very red, then advocating exactly the right amount of dust to make the two indistinguishable would of course seem contrived. This argument does however become difficult to apply if only a small sample of hosts is studied, or if one would like to entertain the possibility that {\it all} quasar host galaxies are young but significantly reddened by dust.

If the hosts of quasars in candidate ejection systems turn out to be very red, it may therefore be necessary to complement the photometry with emission line data to rule out the possibility of dust reddening. A measurement of the $\mathrm{H}\alpha/\mathrm{H}\beta$ ratio using either off-nuclear (\citealt{Hughes et al.}) or on-nuclear spectroscopic techniques (\citealt{Courbin et al.}; \citealt{Jahnke et al. b}) currently feasible for low-redshift objects should allow an estimate of the extinction (see \citealt{Kauffmann et al.} for a discussion on the application of this technique in the host galaxies of AGN), and would also constrain such potential complexities as a quasar redshift different from that of its host galaxy. Although optical reddening at high optical depths may be a poor probe of reddening in colours involving both optical and near-IR filters (\citealt{Witt & Gordon}), this technique should at least provide an indication of whether dust effects are significant or not. Both of these emission lines should display large equivalent widths in the case of quasar-ionized nebulosities and very young stellar population, diminishing the impact of underlying stellar absorption and making a rather low-quality spectrum sufficient for analysis. Narrowband $\mathrm{H}\alpha$ and $\mathrm{H}\beta$ photometry may be an alternative, but could be complicated to interpret due to the unknown velocity fields of the target host galaxies.

\subsection{IMF issues} 
The predicted spectral evolution of aging stellar populations used when deriving the colour criteria of section 3 assumes the stellar IMF to be continuous. This assumption breaks down for very low-mass systems, where the IMF becomes a stochastic distribution due to the small number of stars present. The minimal implementation of the host galaxy test presented here can therefore only be applied to objects at luminosities $M_\mathrm{V} \la -12$ (\citealt{Cervino & Luridiana}). Hence, the proposed test is not suitable for studying the most nearby quasar-galaxy associations, like M82 (\citealt{Burbidge et al.}), where the quasars in the framework of ejection scenarios are suggested to have luminosities below this limit. 

In principle, it could be argued that the strange stellar populations surrounding ejected quasars need not obey anything even remotely resembling a normal IMF. If, for some reason, the upper mass limit of the IMF were substantially lowered, then even a pristine stellar population would appear extremely red, thereby circumventing the proposed test. The Z01 model indicates that in order to reproduce the colours of the \citet{Jahnke et al. a} objects in Fig.~\ref{colfig} with a stellar population younger than 100 Myr, the upper mass limit $M_\mathrm{up}$ of a Salpeter-slope IMF (with lower limit $M_\mathrm{low}=0.08 M_\odot$) would have to drop to $1.5 \la M_\mathrm{up}/M_\odot \la 3$. Although such a stellar population would not evolve into a ``normal'' galaxy (unless the strange IMF is temporary), it would admittedly be very difficult to distinguish from a genuinely old system using broadband photometry. There is however no obvious reason why the hosts of ejected quasars should conspire in this particular way to mimic a normal quasar host galaxy, while intrinsically being so very different.

\subsection{Photoionization modelling}
The presented calculations of the SED of the nebulae photoionized by quasars are admittedly very simplified, assuming spherically symmetric, radiation-bounded and constant-density nebulosities even though large deviations from these conditions could easily be imagined in the vicinity of some strange quasar ejected at a velocity of $\sim 0.1$c and producing matter in its surroundings. The possibilities of ionization by shocks or cosmic rays have moreover not been considered. For this reason, it is quite possible that the colours derived in Section 3 do not reflect the true colour span of quasar-ionized nebulae, and that the colour criteria derived in the framework of the minimal implementation are too strict. Only more extensive ionization modelling can assess this.  

\subsection{Contamination by the active nucleus}
The suggested method is finally prone to some uncertainties which plague the entire field of quasar host research. A particular concern is the possibility that a significant fraction of the light believed to originate in the host galaxy in fact is light from the active nucleus scattered off dust in the interstellar medium of the host. Since the SED of the active nucleus can be very red, such an effect could possibly make a pristine stellar population appear older than it really is. Developing methods which can rule out this possibility is therefore an important task both for quasar host research in general, and for reliable implementation of the proposed quasar ejection scenario test.

\section{Summary}
Although the existence of quasar host galaxies is not acknowledged by the proponents of non-cosmological redshifts as evidence against the notion that some high-redshift quasars may have been ejected from low-redshift galaxies, it has here been argued that such exotic scenarios require the SED of the host galaxies to fulfill a number of criteria, which may be tested with a modest investment of telescope time.  

A minimal implementation of this test has been proposed, in which the SED of quasar hosts at $z<0.7$ may be probed using broadband photometry in two to three filters with a small ($\approx 2.5$ m) telescope equipped with optical and near-IR facilities. No single combination of filters has unfortunately been found to be suitable throughout the entire redshift range. The use of $BRK$ is suggested for $z=0$--0.20, $BVR$ for $z=0.20$--0.35, $VI$ for $z=0.35-0.40$ and $VJ$ for $z=0.40$--0.70. Although the outcome of this minimal implementation could in principle be compromised by the effects of dust reddening or an unusual stellar IMF, the discovery of very blue or absent hosts around the quasars situated in the quasar-galaxy associations claimed to provide the best cases for non-cosmological redshifts would still be an important breakthrough -- as this would indicate that these quasars do in fact stem from a separate population.

\section*{Acknowledgements}
The author would like to thank Nils Bergvall and Eva \"Orndahl for stimulating discussions on topics of quasar host galaxies, galaxy evolution and redshift controversies. Nils Bergvall is furthermore acknowledged for providing the computer code used for convolving synthetic spectra with photometric filter profiles to predict magnitudes at various redshifts. The referee, Alan Stockton, is acknowledged for very useful comments on the manuscript and insightful remarks on the proposed test.

\label{lastpage}
\end{document}